# Graphene-based Tunable Dual-Band Absorbers by Ribbon/Disk Array

Saeedeh Barzegar-Parizi[1],

[1]Electrical Engineering Department, Sirjan University of Technology, Sirjan, Iran
*barzegarparizi@sirjantech.ac.ir

**Abstract**
This article presents two dual-band absorbers based on graphene one dimensional (1D) / two dimensional (2D) patterned arrays in the THz frequencies. In comparison to the previous work, the structure of the proposed device is very simple.  Most dual-band absorbers are based on mixing resonances of resonators with different sizes inside the structure. In this paper, the proposed absorber is based on the arrays of one element. Dual capacitive-inductive nature of the graphene patterned array and the impedance matching condition are used to achieve two resonance bands with perfect absorption by adjusting the graphene properties.

1. **Introduction**

   Graphene, a two-dimensional layer of carbon atoms arranged in a honeycomb lattice, has been introduced as a unique material due to its extraordinary properties of electrical and thermal conductivity [1], high optical transparency, ultrafast electronic transport [2] and controllable plasmonic properties [3]–[5]. The carrier mobility of graphene can be changed by doping methods (chemical or electrostatic doping methods) or by defects creating approaches [6]-[8], and the Fermi level can be modulated in a wide range [9]. Due to the tunability of the carrier mobility of graphene and resulting in its conductivity in the infrared and terahertz ranges, graphene has become one of the most promising materials in many applications such as tunable planar filters [10]-[12], metasurface conformal cloaks [13] and absorbers [14]-[28].
   Recently, the realization of electromagnetic wave (EM) absorbers based on graphene as narrow band [14]-[17], broadband [18]-[23] and multi-band [24]-[28] has attracted the attention of researchers in the THz and mid-infrared (IR) regimes. For the design of the multi-band absorbers, the structures with multi-size patterned elements or asymmetric and complex shapes and the multilayered structures are commonly used. In these structures, the idea is based on mixing resonances of resonators inside the structure. Each resonator produces an absorption mode so that by adjusting two or multi-resonances with specified distances, the multiband absorption occurs. For example, the dual-band absorber realized in [25] is based on the graphene-SiC hybrid system. In [26] and [27], the metamaterial absorbers consist of two sizes of graphene disks and two cross-shaped metallic resonators of different sizes, respectively. In [28], the absorber is based on two layers of graphene as ribbon and sheet.  However, employing the structures with several resonators imposes a complicated fabrication process and therefore designing a simple structure is very important [24].
   In this paper, we present two dual-band absorbers based on the periodic arrays of one simple element. Two structures are investigated through paper, the first structure consists of a 1D array of graphene ribbons (for TM polarization) placed on dielectric spacer terminated by the metal reflector and at the second one, the ribbons are replaced by 2D arrays of graphene disks. The second structure is polarization insensitive for small incident angles. However, both structures are wide angle.
   The graphene patterned arrays show the dual capacitive-inductive nature with a resistive behavior due to the real part of surface conductivity [29]-[31]. When the frequency of the incident wave approaches the resonance frequency, the resonance mode would be excited.  Therefore, the array can be modeled as a surface admittance appeared as an R-L-C series branch at the interface [29]-[32]. One can design a perfect absorber by matching the real part of the input impedance of the structure to the impedance of free space semi-infinite region above the structure where the imaginary part of the input impedance is considered as zero. In the present paper, we use these properties to realize the perfect absorption at two frequencies. Owing to the controllable plasmonic response of graphene, two perfect absorption bands can be realized by adjusting the graphene properties. In comparison to the previous tasks, the most important advantage of the device proposed is its simplicity.

The paper is organized as follows: In Section 2, a dual-band absorber based on graphene ribbons is designed for TM polarization. In Section 3, we realize a polarization insensitive absorber based on the 2D array of disks. Finally, conclusions are drawn in Section 4.

## 2. Dual band absorber based on graphene ribbons

The usual absorbers are made of three layers as the top plasmonic metamaterial, a middle dielectric layer, and a reflective metallic bottom layer [32]-[33]. The top metamaterial layer contains the periodic patterned arrays of one-dimensional and two-dimensional subwavelength elements. The bottom metal film should block light transmission. The excitation of the localized surface plasmon polaritons (LSPP) is the reason for absorbing the light in the most plasmonic metamaterial-based absorbers.

To realization of EM absorber in the low terahertz regime, one can use the graphene-based patterned array. In this section, we survey the graphene ribbon array placed on a dielectric film above a metallic film as shown in Fig. 1(a) to realize the dual-band absorber. The ribbons are assumed with width of $w$ which periodically arranged in the $x$-direction with the period of $L$. The ribbons are considered as infinite in $y$-direction and the surface conductivity of graphene is given by:

$$\sigma_g = \frac{2e^2 k_B T}{\pi \hbar^2} \frac{j}{-\omega + j\tau^{-1}} \ln[2\cosh(E_F / 2k_B T)]$$
$$- \frac{je^2}{4\pi \hbar} \ln\left[\frac{2E_F - \hbar(\omega - j\tau^{-1})}{2E_F + \hbar(\omega - j\tau^{-1})}\right] \quad (1)$$

where $e$ is the electron charge, $E_F$ is the Fermi level, $\hbar$ is the reduced Plank constant, $k_B$ is the Boltzmann constant, $\omega = 2\pi f$ is the angular frequency, $T$ = 300 K is the temperature, and $\tau$ is the relaxation time. The thickness and the refractive index of the dielectric film are considered $h$ and $n_s$, respectively. The structure terminated by a metal layer that is considered as a back reflector.

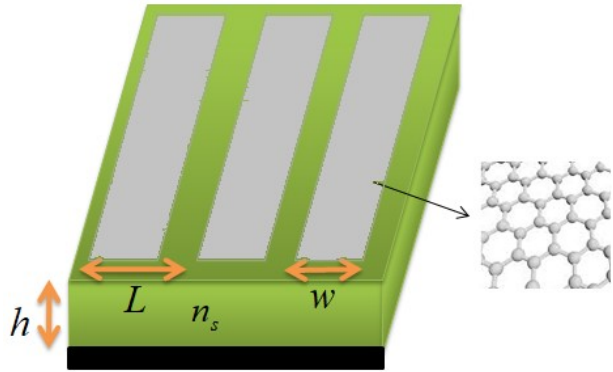

**Fig. 1** Configuration of an EM absorber based on ribbon array placed on top of a dielectric film terminated by the back reflector

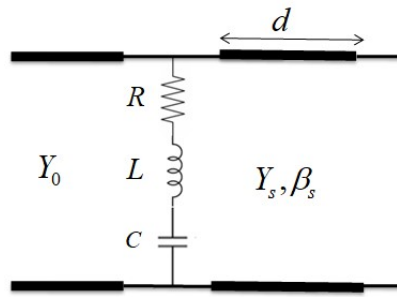

**Fig. 2** The analytical circuit model equivalent to the absorber.

We demonstrate that this device can be employed as a dual-band absorber for TM polarization. The circuit model equivalent to absorber is plotted in Fig. 2 where the ribbon array is modeled as a surface admittance. Hence, the input impedance of the device is written as:

$$Z_{in} = (Y_{sur} - jY_s \cot(\beta_s d))^{-1} \quad (2)$$

where $Y_{sur} = (R + jL\omega + \frac{1}{jc\omega})^{-1}$ is the analytical surface admittance equivalent to ribbon array presented in [29], $Y_s$ and $\beta_s$ are the propagation constant and the admittance of transmission line corresponding to the dielectric slab, respectively. The metallic layer at the bottom of the structure is modeled as the short circuit in the circuit model. In [18], the resonance properties of the graphene ribbons are employed to realize a wideband absorber. In this section, we use these properties to achieve a dual-band absorber. The graphene properties and geometrical parameters of the device are adjusted to obtain two resonant frequencies and the impedance matching conditions for achieving the perfect absorption.

As the first example, consider a device with parameters of $w = 26.8\,\mu m$, $L = 30\,\mu m$ and $h = 24\,\mu m$ where graphene properties are assumed as $\tau = 4.5 \times 10^{-13}\,s$ and $E_f = 0.5\,\text{eV}$. Suppose the refractive index of the dielectric film is $n_s = 3.13$ (Al$_2$O$_3$). Fig. 3 shows the absorption spectra of absorber simulated by HFSS and analyzed by the analytical circuit model. As observed a dual-band absorber can be realized for this device with absorptivity above 99% at 0.52 and 1.4 THz.

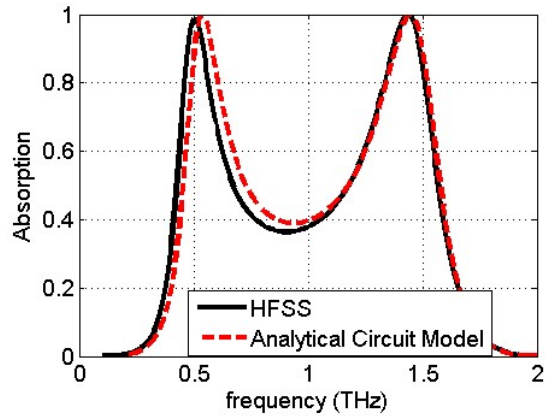

**Fig. 3** Absorption spectra of the absorber based on graphene ribbon array with properties of $w = 26.8\,\mu m$, $L = 30\,\mu m$ and $h = 24\,\mu m$. The properties of graphene appear as $\tau = 4.5 \times 10^{-13}\,s$ and $E_f = 0.5\,\text{eV}$

The real and imaginary parts of the normalized input impedance ($Z_{in}/Z_0$) of the proposed absorber are shown in Fig. 4. $Z_0$ is the impedance of free space semi-infinite region above the structure. It can be seen that the imaginary part of the normalized input impedance is zero at two absorption bands while the real part is close to one. These features represent the impedance matching conditions to obtain the perfect absorption at two resonance frequencies.

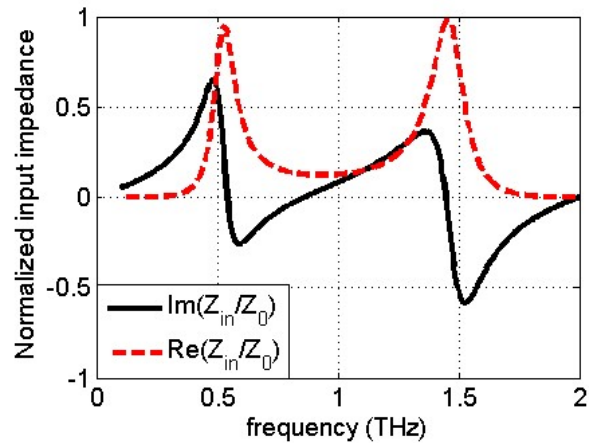

**Fig. 4** Normalized input impedance of the absorber of Fig. 1 with properties presented in Fig. 3.

In the following, we survey the effects of the period, the Fermi level of the graphene and the angle of the incident wave on absorption spectra. In Fig. 5, the absorption spectra are plotted for two different periods. The resonance frequencies are varied by changing the geometrical parameters of the absorber.

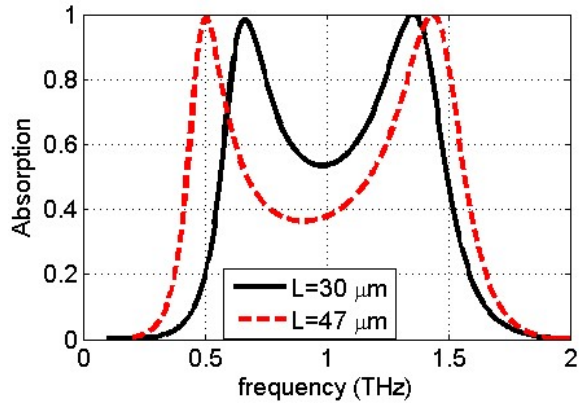

**Fig. 5** Absorption spectra of the absorber with properties presented in Fig. 3 for two different periods.

Now, we survey the Fermi level changes on the absorption bands. Fig. 6 shows the absorption spectra as function of frequency and the Fermi level of the graphene. As observed, owing to the controllable plasmonic response of graphene, the absorption bands can be effectively tuned by the Fermi level of graphene. The Fermi level can be modulated in a wide range by changing the carrier mobility of graphene by doping and defect creating approaches [6]-[8]. Two absorption bands with absorption above 95% can be realized for the Fermi level larger than 0.4 eV. Furthermore, the first resonance frequency is slowly shifted by changing the Fermi level of graphene while the shift of the second frequency is considerable.

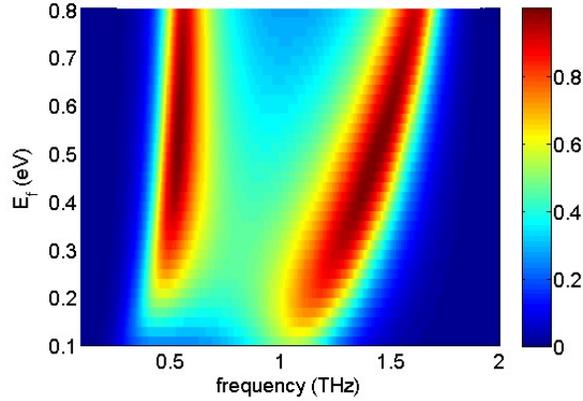

**Fig. 6** Absorption spectra as function of frequency and the Fermi level of the graphene for the absorber with properties presented in Fig.3.

Now, we survey the angle of the incident wave on absorption spectra. The absorption spectra as function of frequency and the incident angle is plotted in Fig. 7. As observed, the proposed absorber has a good performance over a broad range of incident angle (below 50º).

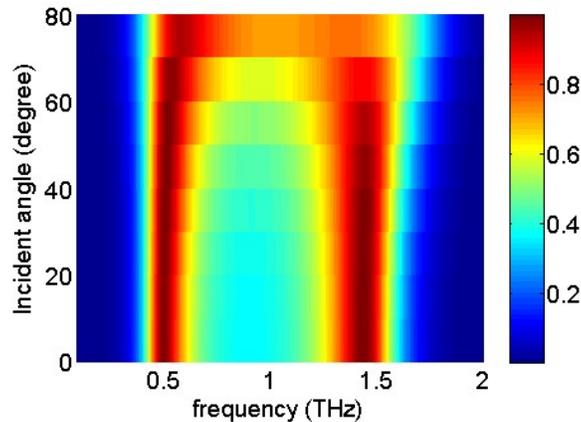

**Fig. 7** Absorption spectra of the absorber as function of incident angle and frequency for TM polarization.

At the end, we demonstrate which the absorption bands can be designed for higher THz frequencies by varying the geometrical and the graphene properties. Fig. 8 shows the absorption spectra for the device with properties $w = 10\,\mu m$, $L = 12\,\mu m$ and $h = 12\,\mu m$ where graphene properties are assumed as $\tau = 2.5\times10^{-13}\,s$ and $E_f = 0.8\,eV$. As observed this device acts as a dual-band absorber with absorption 100% at 1.2 and 2.7 THz. The Fermi level changes are investigated on the absorption spectra as plotted in Fig. 9. For the Fermi level larger than 0.6 eV, the absorption above 95% can be realized for both resonances.

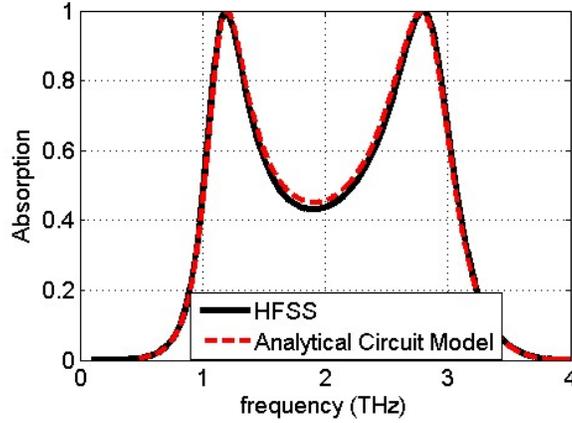

**Fig. 8** Absorption spectra of the absorber based on graphene ribbon array with properties of $w = 10\,\mu m$, $L = 12\,\mu m$ and $h = 12\,\mu m$. The properties of graphene appear as $\tau = 2.5\times10^{-13}\,s$ and $E_f = 0.8\,eV$

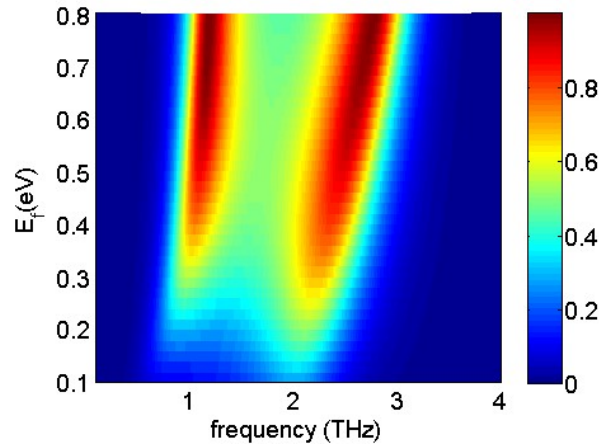

**Fig. 9** Absorption spectra as function of frequency and the Fermi level of the graphene for the absorber with properties presented in Fig. 7.

### 3. Dual band absorber based on graphene Disks

In this section, we realize a dual-band absorber based on graphene disks. This absorber is polarization independent for normal incident waves owing to its symmetric structure. In addition to, we show that this polarization insensitive characteristic remains over a wide range of incident angles.

For designing the absorber, the graphene ribbon array of Fig. 1 is replaced by the 2D array of graphene disks as plotted in Fig. 10. The radius of disks is assumed as $a$. Disks are periodically arranged in the *x*- and *y*- directions with a period $L$. The circuit model equivalent to device can be considered as Fig. 2 where the analytical surface admittance equivalent to disk array presented in [31].

Consider device of Fig. 10 with parameters $a = 24\,\mu m$, $L = 53\,\mu m$ and $h = 24\,\mu m$ where graphene properties are assumed as $\tau = 4.5\times10^{-13}\,s$ and $E_f = 0.7\,eV$. Suppose the refractive index of the dielectric film is $n_s = 3.13$ ($Al_2O_3$). As Fig. 11 shows, a dual-band absorber can be realized for this device.

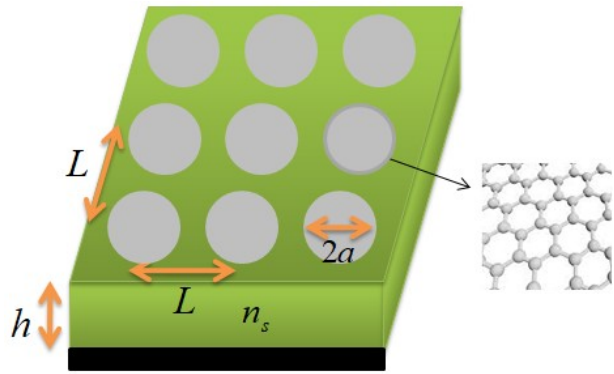

**Fig. 10** The configuration of an EM absorber based on disk array placed on top of a dielectric film terminated by the back reflector

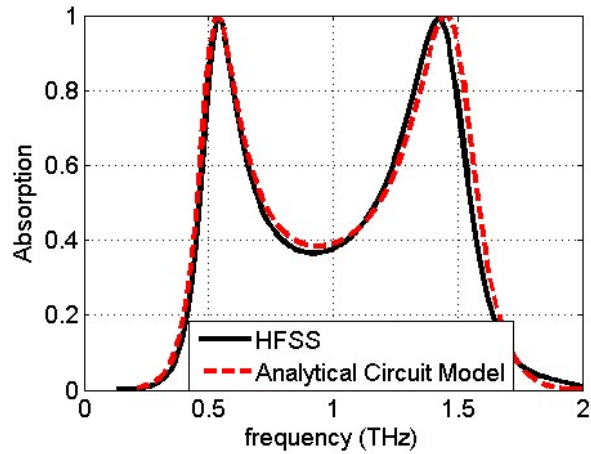

**Fig. 11** Absorption spectra of the graphene-based disk array with properties $a = 24\,\mu m$ and $L = 53\,\mu m$. The properties of graphene appear as $\tau = 4.5\times 10^{-13}\,s$ and $E_f = 0.7\,eV$

Now, we survey the angle of the incident wave on absorption spectra for TE and TM polarizations. Fig. 12 (a) and (b) show the absorption spectra as function of frequency and azimuth incident angle for TE and TM polarizations, respectively. As observed, the proposed absorber has a good performance over a broad range of incident angle for both TM and TE polarizations. In addition to, for the incident angle below $60^0$, there is a substantial overlap between both polarizations; therefore, the absorber is polarization independent at this range. Similar to previous section, the same discussion can be described to survey the effect of the geometrical parameters and graphene properties on the absorption spectra. It is worth to describe that for realization of dual-band absorber in the mid-infrared regime, one can use the metallic patterned array [20]. Since the metals exhibit plasmonic resonance in the mid-infrared and optical ranges.

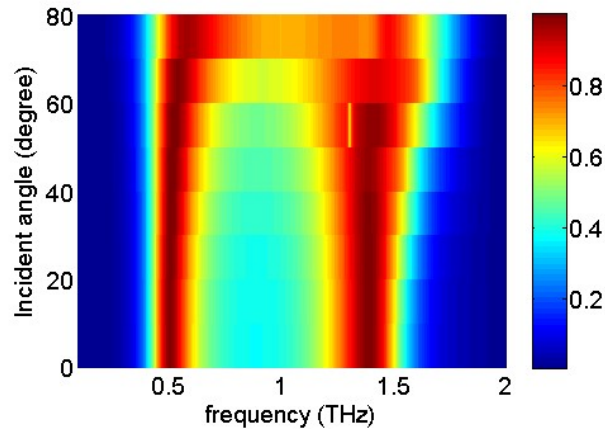

(a)

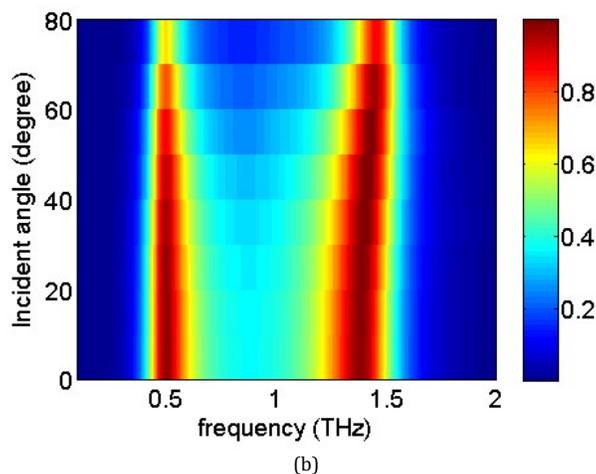
(b)

**Fig. 11** Absorption spectra of the absorber of Fig.9 as function of incident angle and frequency for (a) TM and (b) TE polarization.

## 4. Conclusion

In the current study, two dual-band absorbers have been investigated for the THz frequencies based on graphene patterned arrays. We used 1D array of graghene ribbons in TM case and 2D array of graphene disks to design dual-band absorbers. The Fermi level changes have surveyed on absorption bands. It was demonstrated that two absorption bands with absorvitiy above 95% can be achieved for a wide range of the Fermi levels. Moreover, we showed this device is polarization-insensitive for incident angles below $60^0$ when 2D array of disks was employed to design absorber.